# Fission Thrust sail as booster for high Δv fusion based propulsion

Frederik Ceyssens - Kristof Wouters - Maarten Driesen
*Icarus Interstellar*

## Abstract

The fission thrust sail as booster for nuclear fusion-based rocket propulsion for future starships is studied. Some required aspects of these systems such as neutron moderation and sail regeneration are discussed. First order calculations are used together with Monte Carlo simulations to assess system performance. When the fusion rocket has relatively low efficiency (~30%) in converting fusion fuel to a directed exhaust, adding a fission sail is shown to be beneficial for obtainable delta-v. Also, this type of fission-fusion hybrid propulsion also has the potential to improve acceleration. Other advantages are discussed as well.

## Introduction

The main lesson the rocket equation teaches is the need for high exhaust speed in order to enable the rocket to reach high velocity. Many advanced nuclear propulsion concepts are based on the high speed of the reaction products of nuclear fission and fusion reactions, which are in the order of $10^7$ m/s (3.5%c). These reaction products are then in some way directed out of the rocket before they lose most of their energy by thermalizing collisions with lower energy particles from the reaction mass or with structural parts of the rocket. If this can be achieved, according to the rocket equation a speed Δv of about 10%c is reachable with a single stage rocket with a mass ratio of 20.

Especially for interstellar missions, a high Δv will be of paramount importance. In the Icarus project [1], a choice was made for a deuterium-deuterium fusion rocket, of which the charged reaction products are deflected by a magnetic nozzle and directed backwards into space. Of course, the neutrons cannot be deflected magnetically. The purpose of this paper is to assess if it would make sense to use these neutrons to incite nuclear fission reactions to generate additional propulsion force.

More specifically, the examined case is that of a so-called fission sail or thrust sheet, attached to the spaceship. The fission sail would be out of an inert material, covered on the inner side with a thin film of fissionable material exposed to the neutron flux generated by the fusion reaction in the main engine. The low thickness of the fissionable material film would allow a significant fraction of the fission reaction products to escape at high speed, causing additional propulsive force. The fissionable-elements covered sail concept was originally proposed by Mockel [2], but not explored in combination with a nuclear fusion engine. The concept is shown in figure 1, with some additional features introduced later in this paper. Other non-sail types of fission-fusion hybrids can be found in literature [3-4].

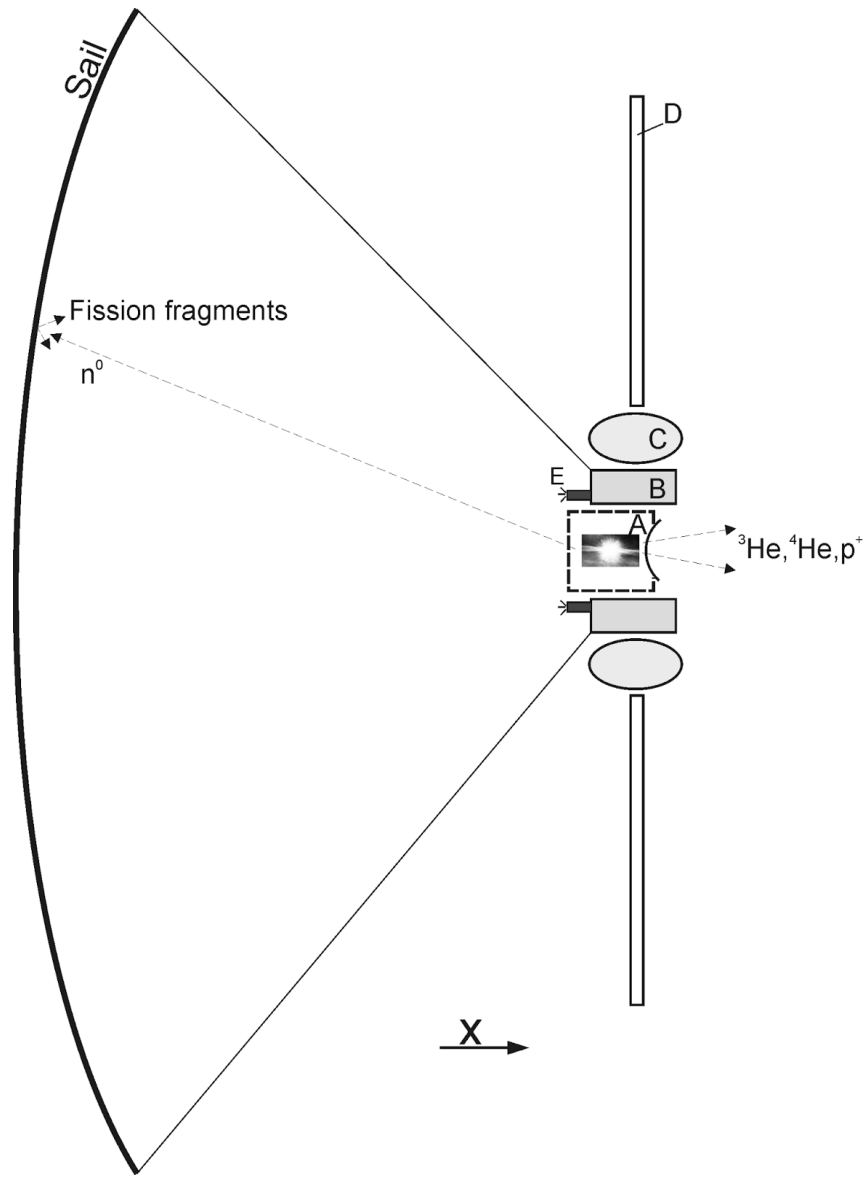

Figure 1: Fission Thrust Sail concept. A: D-D fusion reactor with magnetic nozzle. B: Main ship structure. C: deuterium fuel tank. D: Heat radiator. E: Fission fuel store and evaporation source for sail regeneration. Sail is coated with fissile material on its inner side.

## Basic nuclear reactions

The D-D reaction is as follows:

(50%)  D + D  →  T (1.01 MeV) + $p^+$ (3.02 MeV)
(50%)  D + D  →  $^3$He (0.82 MeV) + $n^0$ (2.45 MeV)

The produced T reacts further:
D + T  →  $^4$He (3.5 MeV) + $n^0$ (14.1 MeV)

The total reaction being:
5D → ³He + ⁴He + 2 n⁰ + p⁺

The properties of the reaction products are displayed in more detail in table 1. In that table, it can also be seen that the average fission product has a speed similar to that of fission fragments. Its rest mass and momentum are much higher, though.

|  | Ek [MeV] | speed [%c] | rest mass [kg] | momentum [kg m/s] |
| --- | --- | --- | --- | --- |
| p⁺ | 3.02 | 5.7 | 1.67E-27 | 2.84E-20 |
| ⁴He | 3.5 | 3.1 | 6.68E-27 | 6.11E-20 |
| ³He | 0.82 | 1.7 | 5.01E-27 | 2.56E-20 |
| n⁰ | 2.45 | 5.1 | 1.67E-27 | 2.56E-20 |
| n⁰ | 14.1 | 12.2 | 1.67E-27 | 6.09E-20 |
| aver. ²³⁹Pu fission product | 175.8 | 4.0 | 1.97E-25 | 2.36E-18 |

Table 1: reaction products of D-D fusion reaction. Relativistic formulas are used calculating speed and momentum out of Ek and rest mass. The weight-averaged speed of charged fusion products at the time of generation is 0.029c. The weight-averaged speed of all particles, of which the neutron speeds are multiplied by $2/\pi^2$ to account for their uniform distribution over all angles and the fact that only half of the neutrons goes in the right direction, is $v_F$ = 0.0265c. Their average mass $m_F$ is $3.34 \cdot 10^{-27}$ kg.

## Table of symbols

- $\sigma$ — microscopic cross section [barn=1e-28 m²]
- $\sigma_{D,c}$ — Deuterium neutron capture cross section
- $\sigma_{D,s}$ — Deuterium neutron scatter cross section
- $r_\sigma$ — (fractional) number of fission reactions per absorbed neutron
- N — atomic density [nb of atoms / cm³]
- $v_{fiss}$ — starting velocity of fission fragments [m/s]
- $r_m$ — range of charged fission fragments in solid matter [m]
- $r_{mass}$ — mass increase factor when adding sail
- $m_{fiss}$ — average mass of a fission fragment [kg]
- $m_F$ — average mass of a fusion product
- $m_{fissile}$ — atomic mass of fissile fuel [Da]
- $m_{fissile}$ — atomic mass of deuterium [Da]
- $M_{empty}$ — mass of pure fusion rocket without fuel [kg]
- $M_D$ — mass of deuterium fuel at start [kg]
- $M_{fiss}$ — mass of fissile fuel at start [kg]
- $\eta_F$ — fusion efficiency
- $\eta_n$ — neutron efficiency
- $v_F$ — weight-averaged speed of fusion products when generated [m/s]

## Neutron absorption and fission product range

The absorption of neutrons is modelled by the microscopic cross section σ. For the relevant materials, the most important effect after neutron absorption is nuclear fission. Nuclear interactions other than fission and neutron capture such as scattering are neglected as the respective cross sections are orders of magnitude smaller for the relevant materials and neutron energies, as will become clear later. The fraction $r_\sigma$ represents the number of fission reactions initiated per absorbed neutron. For an impinging neutron beam with intensity $I_0$ at the surface, the remaining intensity Ix at a depth x is then:

$$I_x = I_0 e^{-N \cdot \sigma \cdot x} \qquad (1)$$

With N the atomic density of the medium (for uranium, N = 0.048 $10^{24}$ / cm³.).

The fission products are charged particles and are, alas, stopped relatively fast compared to neutrons. The approximate relationship between speed *v* and distance *r* for charged particles is:

$$\frac{v}{v_{fiss}} = 1 - \frac{r}{r_m} \qquad (2)$$

with $v_{fiss}$ the initial speed of the charged particle (about 1.2e7 m/s) [5] and $r_m$ the range. According to [5], the range $r_m$ of fission fragments in U is 0.66 e-5 m.

Thus, in order not to waste too much fissionable material on the sail, it is necessary that a most of the impinging neutrons are absorbed within the first few micrometers of the sail. If this turns out to be impossible as will become clear soon, it is advisable to make the fissionable material layer not thicker than a few micrometers anyway: a significant fraction of the neutrons will not be causing fission then, but at least no fissionable material (reaction mass) is wasted.

In figure 2, the absorption depth (1/Nσ) of neutrons as a function of cross section is displayed. It can be seen that a cross section of over 1000 barn at the very least and preferable over 10000 barn is required for a significant fraction of neutrons to be absorbed in a layer of only a few micrometers thin.

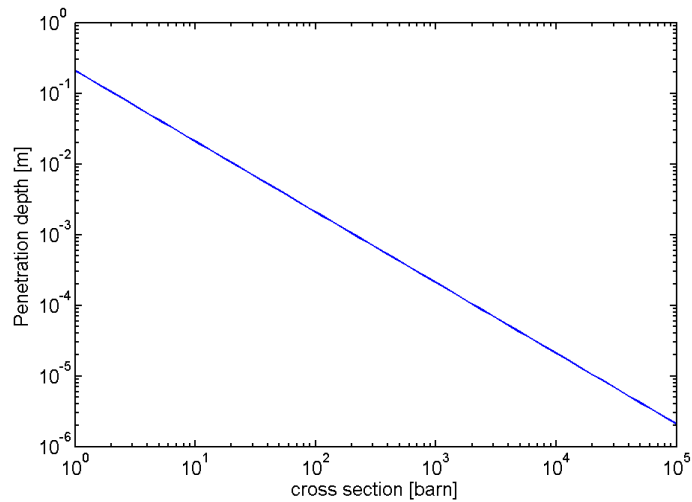

Figure 2: absorption depth (depth at which a fraction of 1/e of impinging neutrons is left) in uranium as a function of cross section

Such high cross sections can be obtained when using fissile materials ($^{235}$U, $^{239}$Pu,...) as illustrated in figure 3. Also, these materials also show negligible neutron scattering at these energies; their main mode of neutron interaction is fission. In the rest of this paper, we will therefore assume $^{235}$U or $^{239}$Pu as fissile material covering the sail. Other fissile materials are too scarce to be considered. Furthermore, a second necessary condition is the use of very low-energy (cold) neutrons, having energies in the meV range. Thus, an important part of the propulsion system design will be a device to slow the fusion-generated fast neutrons down to low energy.

In current nuclear technology, a layer of cold deuterium or heavy water is used for this, which is convenient as it is present as nuclear fusion fuel anyway. An important parameter to consider is the efficiency $\eta_{n,mod}$, i.e. the amount of cold neutrons exiting the device for every fast neutron entering it.

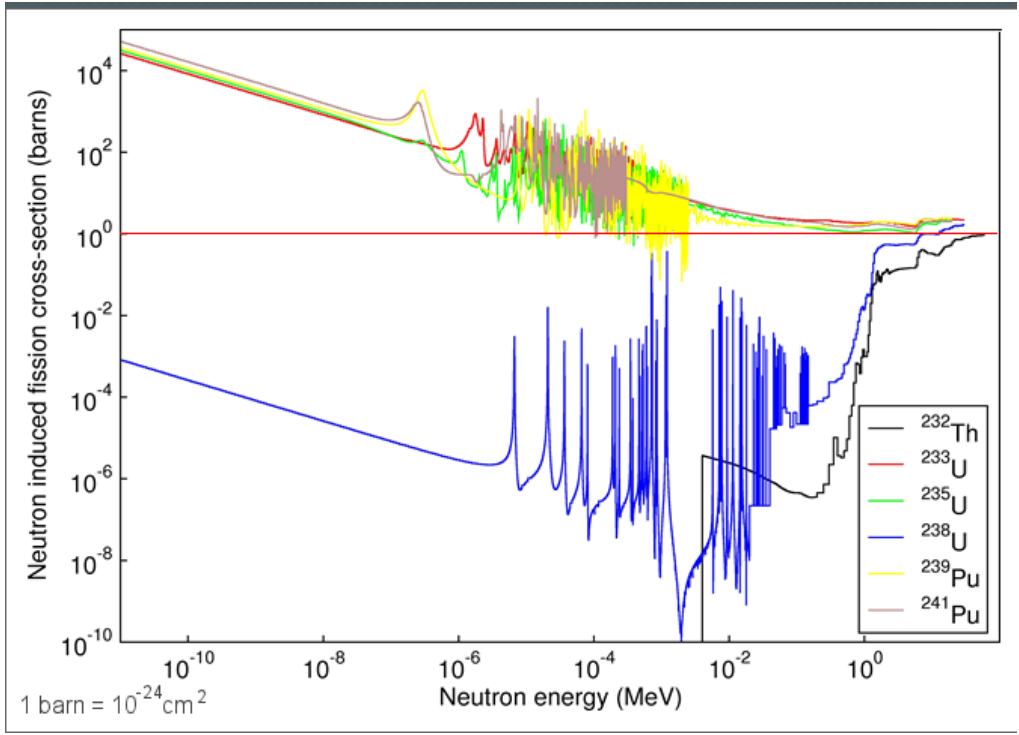

Figure 3: Fission cross-sections as a function of neutron energy [6]

# Exhaust speed calculation

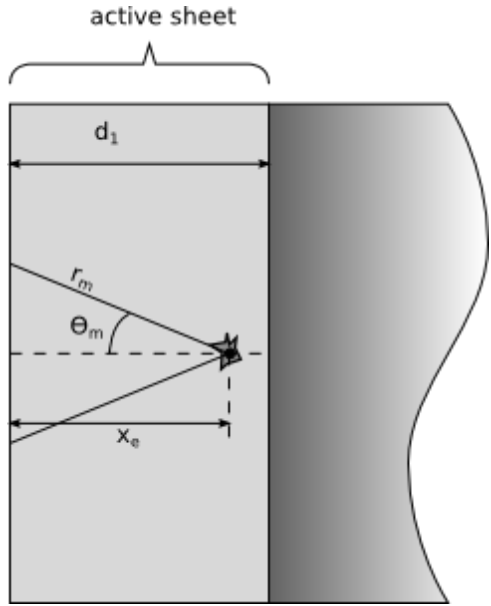

Figure 4: a schematic cross section view of the thrust sheet.

We will now calculate generated thrust per impinging neutron, with the fission cross section as most important parameter.

For fission products emitted at a depth $x_e$ ($x_e <= r_m$), the maximum angle $\theta_m$ (with respect to the normal to the sail) at which the particles will still reach the surface will be:

$$\theta_m = arcos(\frac{x_e}{r_m}) \qquad (3)$$

We'll now calculate $<v_{surf}>$, the average speed in the x direction *at the surface of the sail* of a particle emitted at speed $v_e$ at depth $x_e$, assuming semi-spherical isotropic distribution. Note that two particles are always emitted in opposite directions. The second particle is always stopped by the sail, and is not included in the calculation below.

Integrating the x-axis projection of v over the relevant surface (the intersection of the cone with apex angle $\theta_m$ and the sail surface) and dividing by that surface area yields the average particle speed $v_{x,av}(x)$ at the surface of the sail, for particles emitted at depth x. A factor $2\pi(1-cos(\theta_m))$ needs to be added to account for the particles outside of the cone, those that do not make it to the surface.

$$v_{x,av}(x) = \frac{\oint_S v_{fiss}(1-\frac{r}{rm}) \, sin(\theta)cos(\varphi) \, dS}{Area(S)} 2\pi(1 - cos(\theta_m)) \qquad (4)$$

with $\theta$ and $\varphi$ as in the normal convention for spherical coordinates and S the base of a cone with height x and apex angle $\theta_m$.

To obtain the average speed at the surface in the x axis direction $<v_{surf}>$, (4) is averaged with a weighting factor from equation (1).

$$<v_{surf}> = \frac{\int_0^{min(d_1,r_m)} v_{x,av}(x)\, e^{-N\cdot\sigma\cdot x}\, dx}{\int_0^{min(d_1,r_m)} e^{-N\cdot\sigma\cdot x}\, dx} \qquad (5)$$

As no analytical solution was found, these integrals were solved numerically using a Monte Carlo method. A random number generator was used to generate particles emitted at depth x. Using (2) and (3), the speed of those particles at the surface was calculated. An average was taken over 10000 particles. A weighted average with weighing factor (1) was then applied over a depth range [0 $d_1$]. The results are plotted in figure 5.

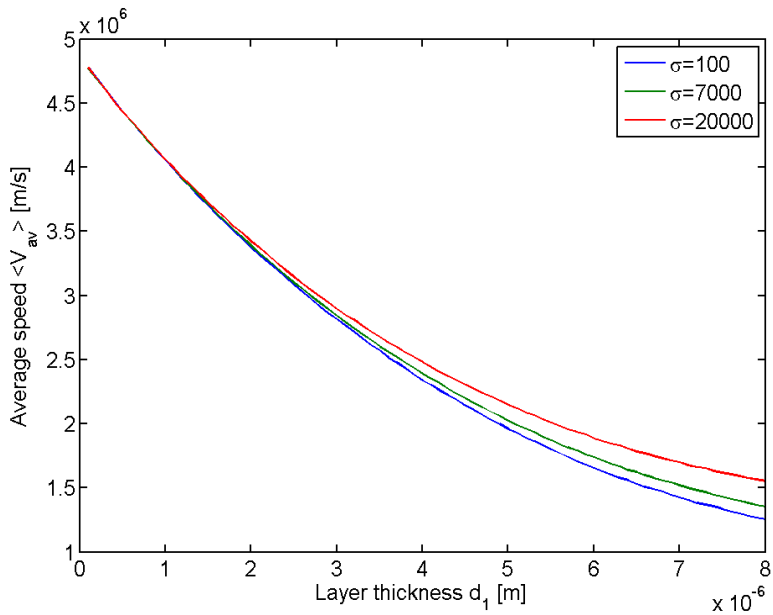

Figure 5: average x-axis speed of fission products generated in a layer of thickness $d_1$.

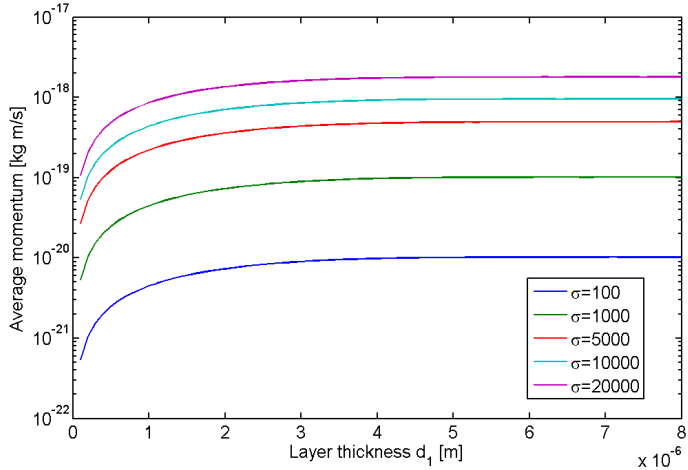

Figure 6: average x-axis momentum produced per neutron impinging on a $^{235}$U sail, with fission cross section σ as parameter.

The average x-axis momentum of a fission fragment at the surface can now also be calculated, based on an average fission particle mass $m_{fiss}$ of 118 Da.

At this, point, it must be noted that not all absorbed neutrons cause fission. We therefore introduce a factor $r_\sigma$ to account for the fact that sometimes no fission reaction is initiated by neutron absorption, but an isotope 1 unit higher in atomic mass is generated instead. For $^{235}$U, $r_\sigma$ equals 0.855 [7]. Upon neutron capture without fission, the low cross section isotope $U^{236}$ is generated, which does not react further (see section: sail regeneration). For $PU^{239}$, 23.5% of captured neutrons generate $^{240}$Pu, which then captures a second slow neutron to transmute to $^{241}$Pu, which has a fission cross section 35% higher than that of $^{239}$Pu and a 73% fission probability. This translates into a need for 1.47 neutrons for 0.93 fission reactions, or $r_\sigma$ = 0.6337. Therefore, the performance of $^{235}$U can be expected to be superior even though the fission cross section of $^{239}$Pu is about 20% higher [8].

Furthermore, the average momentum produced per neutron impinging on the sail as a function of fission cross section (depending on neutron energy) and layer thickness can be calculated, by multiplying with the fraction $r_\sigma(1 - e^{-N \cdot \sigma \cdot d_1})$ of impinging neutrons used. The results are plotted in figure 6. It can be seen that there is almost no rise in average momentum from a layer thickness of 3 micrometers on. For that layer thickness, the average particle speed at the surface is 0.96%c. This speed is only weakly dependent on σ. With higher $d_1$ there will only be extra consumption of fissile material without any benefit to thrust.

### Performance calculations

We have already stated that a significant fraction of neutrons from the fusion reaction will get lost without having incited a fission reaction. We therefore define a *neutron efficiency* $\eta_n$ as the fraction of fusion-generated neutrons that actually hits the sail. The *neutron efficiency* $\eta_n$ can be subdivided in geometrical effects (neutrons hitting the reactor wall, the sail only being located on one side of the ship etc.) and absorption in the neutron moderator between the fusion reactor and the sail:

$$\eta_n = \eta_{n,geom} \cdot \eta_{n,mod} \qquad (6)$$

A reasonable estimation of the obtainable value for $\eta_{n,geom}$ is 0.3, or 0.5 if neutron reflectors are used to deflect more neutrons to the sail. For $\eta_{n,mod}$, we estimate 0.57 as a realistic value discussed in the appendix. The loss due to non-complete absorption in the sail is not factored in, and will be included separately in the equations.

Likewise, not all charged fusion particles generated will be expelled from the rocket along the axis of flight before significant thermalisation or other loss has occurred. This is not always recognised in work about future starship propulsion, where all too easy the full speed of the fission fragments is put in the rocket equation. We therefore define also a fusion particle emission efficiency $\eta_F$. $\eta_F$ is the fraction of the momentum of the fusion products that can potentially be generated with a certain quantity of fuel that is actually converted into thrust. It not only models the efficiency of the magnetic nozzle, but also include all other effects that drain the energy from the exhaust elsewhere, such as the conversion of some of the energy into electricity to power the ship or the fractional burnup of fusion fuel.

Another factor to consider is the additional system mass (i.e. excluding fuel) that adding a fission thrust sail will impart on the ship. This mass arises from e.g. the fact that the fuel tanks have to be placed at a larger distance in order to make the reactor more open for neutrons to escape and reach the sail, the additional auxiliary system, the mass of the inert part of the sail, etc. Due to the simplicity of the thrust sail system, the additional mass is expected to be low or even negative, if a less powerful fusion reactor is needed for the same acceleration (also entailing less powerful heat radiators), and the need for less deuterium fuel with therefore a smaller cryogenic storage system for this low-density liquid. Furthermore, as already stated the neutron moderator which can be expected to be one of the heavier components of the thrust sail system, can use the same deuterium as is present as fusion fuel anyway. The relative change in dry mass caused by adding the fission sail is modeled as mass factor $r_{mass}$.

$\eta_n$, $\eta_F$ and the additional mass factor $r_{mass}$ will determine whether or not the addition of the thrust sail is sensible.

The delta-v for a pure fusion based propulsion system is simply given by the rocket equation:

$$\Delta v_F = \eta_F \cdot v_F \cdot ln\left(\frac{M_D + M_{empty}}{M_{empty}}\right) \quad (7)$$

with $M_D$ the mass of the deuterium fuel and $M_{empty}$ the mass of the rocket with empty fuel tanks.

To calculate the contribution of adding a fission thrust sail, one must consider that:
- for every 5 fusion product particles, only 2 neutrons are produced
- the *neutron efficiency* $\eta_n$ and the factor $r_\sigma(1 - e^{-N \cdot \sigma \cdot d_1})$ further limit the amount of used neutrons
- one of the two fission products always gets stuck in the sail, which was not factored in so far.
- a mass weighted average should be taken of the speed of the different particles
- adding the sail causes a change in dry mass as discussed above, modeled by the mass factor $r_{mass}$.

Plugging this in the rocket equation leads to a delta-v for a propulsion system with added sail of:

$$\Delta v_{sail} = \frac{2/5 \, m_{fiss} \, \eta_n \, (1-e^{-N\sigma d_1}) r_\sigma <v_{surf}> + m_F \, \eta_F \, v_F}{4/5 m_{fiss} \, \eta_n \, (1-e^{-N\sigma d_1}) r_\sigma + m_F} ln\left(\frac{M_{fiss} + M_D + r_{mass} \cdot M_{empty}}{r_{mass} \cdot M_{empty}}\right) \quad (8)$$

Note that there is no $\eta_F$ in the denominator as this is a velocity (or impulse) ratio and not a mass ratio. The change in initial acceleration of the rocket when adding the fission sail is now calculated.

The total mass of fissile elements $M_{fiss}$ that has to be carried next to the deuterium fuel mass $M_D$ is now calculated. For every deuterium atom used as fuel, the number of reacting fissionable elements is $2/5 \, \eta_n \, (1 - e^{-N \cdot \sigma \cdot d_1})$.

Therefore,

$$M_{fiss} = 2/5 \, \eta_n \, (1 - e^{-N \cdot \sigma \cdot d_1}) \frac{m_{fissile}}{m_D} M_D \qquad (9)$$

with the atom mass ratio $\frac{m_{fissile}}{m_D}$ equal to 119.5 for $^{239}$Pu and deuterium and 117.5 for $^{235}$U and deuterium.

The weight change of the empty rocket when adding the sail system is modelled again with the mass factor $r_{mass}$.

The rocket's acceleration at start scales with a factor $r_a$:

$$r_a = \frac{a_{fission}}{a_F} = \frac{I_{fiss}}{M_{tot,fiss}} \frac{M_{tot,F}}{I_F} \qquad (10)$$

with $I_{fiss}$ and $I_F$ the sum of the momenta of the released particles *per deuteron* in case of an added fission sail and pure fusion respectively, and $M_{tot,fiss}$ and $M_{tot,F}$ the total ship mass in the two cases.
From Table 1:

$$I_F = \eta_F \sum(fusion\ particle\ moments) / 5 \qquad (11)$$

Furthermore,

$$I_{fiss} = 2/5 \, \eta_n \, (1 - e^{-N\sigma d_1}) r_\sigma \, m_{fiss} <v_{surf}> + I_F \qquad (12)$$

The fraction of total masses of a fission sail-boosted and a pure-fusion rocket is:

$$\frac{M_{tot,fiss}}{M_{tot,F}} = \frac{M_D + M_{fiss} + r_{mass} \cdot M_{empty}}{M_D + M_{empty}} \qquad (13)$$

In the following section, a performance estimation is made for a series of typical parameters. For the pure fusion case a ship with $M_{empty}$ equal to 1000 tonnes is assumed, with a mass ratio $(M_D + M_{empty}) / M_{empty}$ of 20. In order to assess the performance of the fission sail, the equations above are solved for the following parameters:

| | |
|---|---|
| $r_{mass}$ | 0.8, 1, 1.2 |
| ($\sigma$ [barn], $\eta_n$) | (3000, 0.288), (3000, 0.05), (3600, 0,288) |
| $d_1$ [m] | $3 \cdot 10^{-6}$ |

Table 2: parameters used for calculating figures 7-8

The lower cross sections are not evaluated with lower $\eta_n$ as they occur with thermal neutrons for which $\eta_n$ is higher. $\eta_F$ is taken as variable and swept between 0 and 1, as no concrete data about its likely value is known to us at this time.

The results are plotted in figures 7 and 8. The most important conclusion is that, in a range of for reasonable values for the fission sail parameters and assuming a neutral effect of adding the sail on the ships dry mass ($r_{mass}$=1), adding the system would make sense when the efficiency of the fusion system $\eta_F$ is below the range [0.29 0.34]. Below that range, the improvement offered can be several hundreds of percents. Acceleration gets significantly below an $\eta_F$ of about 0.5. For an $\eta_F$ below 0.1, improvement can be a factor 4 or more, significantly influencing mission design. Investigating the effect of $r_{mass}$, even if there is a 20% dry mass penalty on adding the sail ($r_{mass}$=1.2), the fission sail is still beneficial for $\eta_F$ below 0.22. For a $r_{mass}$ below 1, the advantage of the sail is significant even at high fusion efficiencies.

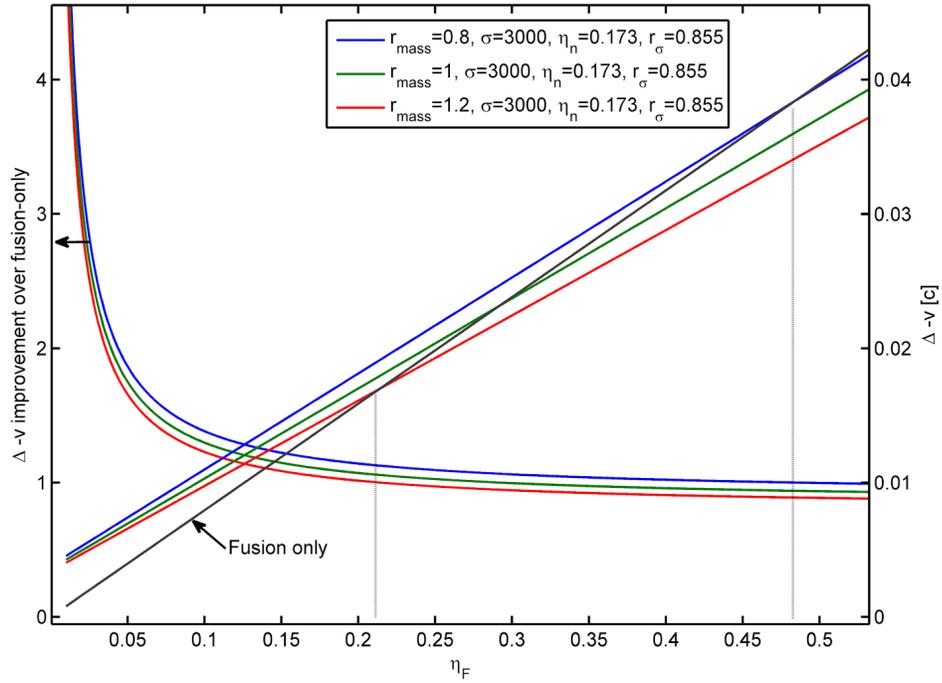

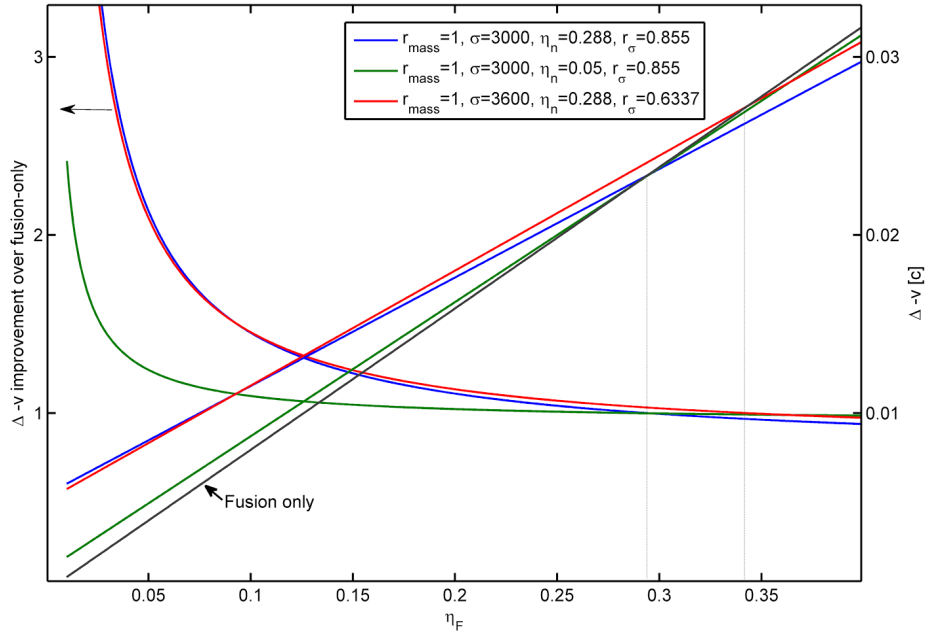

Figure 7: Calculated delta-v improvement factors and absolute delta-vs for a number of cases. The cases with σ=3000 are for $^{235}$U, when σ equals 3600 the fissile material used is $^{239}$Pu.

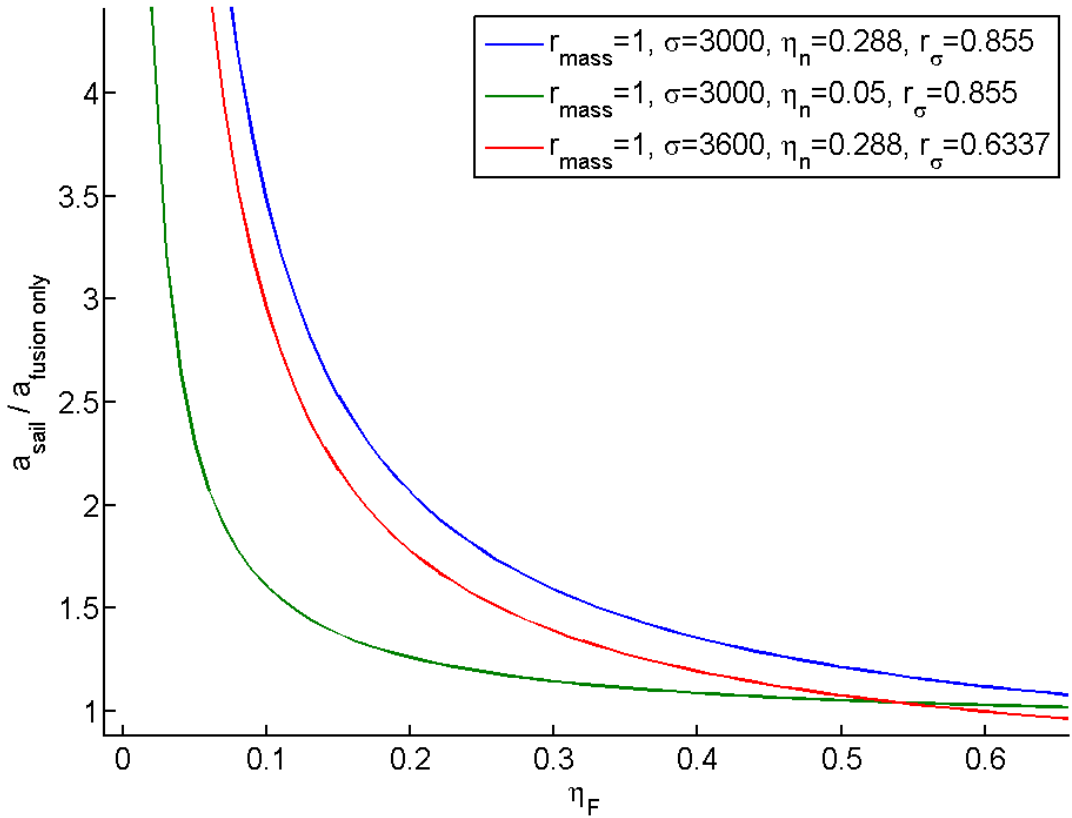

Figure 8: acceleration improvement offered by addition of fission sail.

## Sail regeneration

The low thickness of fissile material allowed on the sail means the layer will be used up quickly, and the thickness will soon diverge from the designed optimal thickness. Also, fission products will accumulate in the sail. It is possible though to dynamically keep the layer thickness at the optimal value by projecting new material from the ship, e.g. using an thermal evaporation source located on the main structure of the ship.

This periodic replenishment of the fission material layer can be preceded by a step in which the sail is heated, boiling off the trapped fission products still present. For this reason we will assume the fission products generated do not contribute to $M_{empty}$. Of course, in that case the fission sail's structural material should then have a higher melting point than that of common fission products. Also, do note that a 500 by 500 meter sail coated with 3 μm of uranium already weighs 14 tons. Thus, these coating cycles will be relatively coarsely spaced in time. Another advantage of employing sail regeneration is that the radioactive fissile layer can be applied to the sail in space, after ship assembly, thus contributing to safety. As a side note, a further advantage of the sail would be a dual use as micrometeorite shield.

## Cost of thrust sail

The current US Department of Energy sales price of Pu-239 for research quantities is about 10000 USD per gram [9], which would amount for a 100 ton sail to one trillion dollars. In the past, when nuclear weapons were mass manufactured, the actual fabrication cost could well be at least an order of magnitude lower. Existing stock could also be used: as of the end of 2012, Russia's fissile material stock is estimated to include about 128 tonnes of weapons-grade plutonium and 695 tonnes of highly-enriched uranium [10]. As the raw material price (uranium oxide) out of which plutonium is bred is only in the order of 50 US$ per kg, this cost can most probably be reduced by several orders of magnitude by process upscaling, if it is decided to develop a thrust sheet based starship for which much higher quantities are needed than for weapons fabrication. Nevertheless, the cost of deuterium, the prefered fuel for pure fusion based propulsion, is about 5 US$ per gram [11] and thus much more affordable.

## Discussion and conclusion

The concept of a fissile element coated sail as a booster for nuclear fusion based rocket ships was introduced and studied. Whether or not adding the sail makes sense to optimize delta-v depends on the efficiency of the fusion based rocket, which is unknown at this moment. According to the calculations presented, the efficiency below which a fission sail makes sense is in the range of 29% to 34% for $r_{mass}$ = 1.

Besides this, there are other advantages offered by the fission sail booster. As the acceleration increases significantly, a mission designed for a certain acceleration could do with a much smaller fusion reactor, and thus smaller heat radiators and cryogenic fuel tanks, improving the mass ratio. Also, the sail can have a double purpose as a micrometeorite shield.

Compared to a laser or microwave-driven sail, the sail is orders factors of magnitude smaller which is a considerable advantage when it comes to construction. Of course, also no external drivers are needed.

Some factors were not covered in the preliminary calculations presented and are left for later work. These include more detailed moderator calculations and applying the calculated neutron energy spectrum to fission product generation instead of using the median, the effect of neutron reflection by the inert part of the sail and a more detailed system design. Also, the behavior of the fissile material layer under neutron bombardment should be studied. Due to lack of data, the model used in his work has significant simplifying assumptions. E.g., it assumes constant properties of the fissile layer on the sail. Also, fission fragments could knock fissile atoms out of the sail before they can react with neutrons, lowering delta-v.

The presented concept was a one-stage design. Moving to a two or more stage design is of course still possible to increase delta-v.

The concept is testable with current technology and a limited budget: a test setup could comprise a sheet a few cm² in area on a cantilever for optical measurement and an accelerator-based neutron. Accelerator-based cold neutron sources with a neutron flux density of $5 \cdot 10^{10}$ cm$^{-2}$ s$^{-1}$ are available [12], yielding nanonewtons of thrust per cm².

All together, the preliminary calculations presented show enough promise to justify further research into this propulsion technology.

___________________________________________

## Appendix: Neutron moderation calculation

The following data are used:

| | |
|---|---|
| Deuterium scatter crossection $\sigma_{D,s}$ [16] | 1.33 - 4.5 barn |
| Deuterium capture crossection $\sigma_{D,c}$ [14] | 0.0000519 barn |
| Liquid deuterium atom density [cm$^{-3}$] | 4.25·10$^{22}$ |
| neutron speed reduction factor per collision μ | 2/3A$_D$ = 1/3 |

Table 3: nuclear diffusion data

From this the macroscopic scattering and capture cross sections are calculated: $\Sigma_{D,S}$ = 0.17 cm$^{-1}$ and $\Sigma_{D,c}$ = 1.27 10$^{-5}$ cm$^{-1}$. The mean free paths of those phenomena are the inverse of the cross sections. The *transport mean free path* $\lambda_{tr}$ can then be calculated [13]:

$$\lambda_{tr} = 1/\Sigma_{tr} = 1/(\Sigma_{D,S}(1-\mu)) \qquad (14)$$

which leads to an approximate diffusion constant D= $\lambda_{tr}$/3.

The diffusion length L is then:

$$L = \sqrt{D/\Sigma_{D,c}} = 480 \text{ cm} \qquad (15)$$

To reduce the neutron from MeV to meV speeds, about 20 collisions are needed. The average total path traveled is thus 20/$\Sigma_{D,S}$ =117.6 cm. This causes a loss of $(1-e^{-117/L}) = 21\%$, i.e. an η$_{n, mod}$ of 79%.

For a more accurate estimation of moderator performance which included neutron backscattering out of the moderator, a Monte Carlo simulation was employed. Main simplifying assumptions employed were:
1. the use of energy-dependent scatter cross sections from [13]
2. an exponential distribution of travel distance between collisions with factor $\Sigma_{D,S}$
3. a uniform distribution of scattering angle in an interval [-2/3π,2/3π]
4. a reduction of speed at every collision by a factor μ=⅓ until thermal equilibrium is reached, thereafter speed is constant.
5. an exponential distribution of neutron loss distance with factor $\Sigma_{D,c}$
6. neutron speeds according to table 1.

Note that the interval in distribution used in assumption (3) was observed not influence efficiency for more than 3% when changed from [-1/3π,1/3π] to [-2/3π,2/3π]. The needed moderator thickness did almost double though. Therefore, we have some confidence in the employed method when it comes to efficiency calculations. For final design, a more careful study is required.

The simulation results are plotted in figure 9. It can be seen that in order to reach a median neutron energy in the few meV range corresponding to a $^{235}$U fission cross section of about 3000 [15], a liquid deuterium moderator layer of 0.49 meter thickness is needed. The efficiency $\eta_{n,\,mod}$ is then 57.7%.

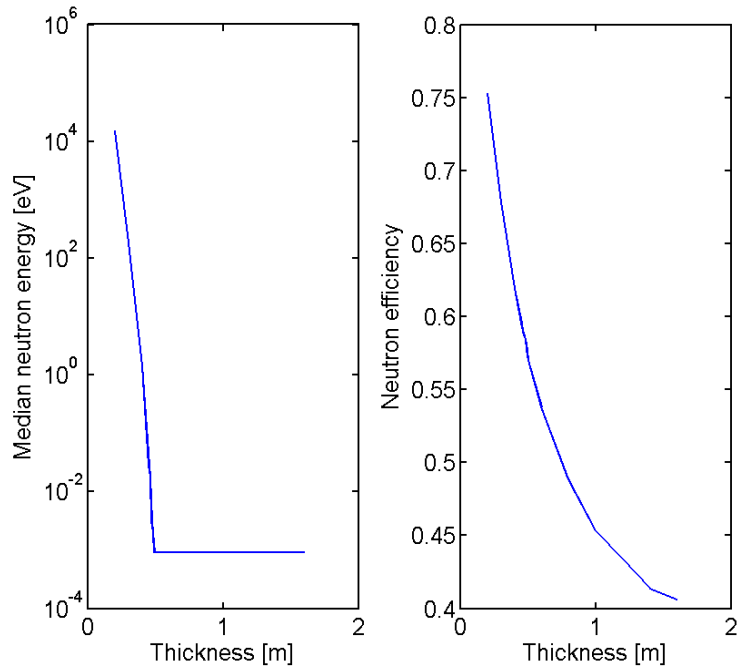

Figure 9: Monte Carlo simulation results, showing median neutron energy and efficiency versus moderator thickness.